\documentclass[11pt, letterpaper]{article}
\usepackage[utf8x]{inputenc}
\usepackage{amsmath}
\usepackage{amsthm}
\usepackage{physics}
\usepackage[margin=1in, bottom=0.75in, top=0.75in]{geometry}
\usepackage{amsfonts}
\usepackage[T1]{fontenc}
\usepackage{enumitem}
\usepackage{blkarray}
\usepackage{fancyhdr}
\usepackage{tikz}
\usepackage{circuitikz}
\usepackage{caption}
\usepackage{float}
\usepackage {mathrsfs}
\usepackage[makeroom]{cancel}
\usetikzlibrary{arrows, shapes, decorations.pathmorphing,backgrounds,positioning,fit,petri}
\usepackage{bbm}
\usetikzlibrary{arrows.meta}
\usepackage{polynom}
\usepackage[pagebackref, colorlinks,citecolor=blue,linkcolor=orange,urlcolor=cyan,bookmarks=true]{hyperref}
\usepackage{graphicx}
\usepackage{subcaption}
\hypersetup{
	colorlinks=true,
	urlcolor=cyan
}
\usepackage{listings}   
\usepackage{multirow} 
\usepackage{hyperref}
\usepackage{caption}
\usepackage{amsthm}
\usepackage{cleveref}
\usepackage{authblk}

\theoremstyle{fact}
\newtheorem{fact}{Fact}[section]

\theoremstyle{theorem}
\newtheorem{theorem}{Theorem}[section]

\theoremstyle{definition}
\newtheorem{definition}{Definition}[section]

\theoremstyle{corollary}

\theoremstyle{conjecture}

\theoremstyle{lemma}
\newtheorem{lemma}[theorem]{Lemma}

\lstdefinestyle{shared}
{
	numbers=left,
	numbersep=1em,
	numberstyle=\tiny\color{red}\noaccsupp,
	frame=single,
	framesep=\fboxsep,
	framerule=\fboxrule,
	rulecolor=\color{red},
	xleftmargin=\dimexpr\fboxsep+\fboxrule\relax,
	xrightmargin=\dimexpr\fboxsep+\fboxrule\relax,
	breaklines=true,
	tabsize=2,
	columns=flexible,
}
\usepackage{titlesec}
\titlespacing*{\section}
{0pt}{0.8em}{0.8em}
\titlespacing*{\subsection}
{0pt}{0.8em}{0.8em}

\usepackage[linesnumbered,lined,ruled]{algorithm2e}
\usetikzlibrary{calc,patterns,angles,quotes}

\newtheorem{innercustomgeneric}{\customgenericname}
\providecommand{\customgenericname}{}
\newcommand{\newcustomtheorem}[2]{%
  \newenvironment{#1}[1]
  {%
   \ifdefined\crefalias\crefalias{innercustomgeneric}{#2}\fi
   \renewcommand\customgenericname{#2}%
   \renewcommand\theinnercustomgeneric{##1}%
   \innercustomgeneric
  }
  {\endinnercustomgeneric}%
  \ifdefined\crefname\crefname{#2}{#2}{#2s}\fi
}

\newcustomtheorem{customthm}{Theorem}

\title{Lower Bounds on Relative Error Quantum Compression and Classical Shadows}
		
\author{Kaushik Sankar\thanks{Department of Computer Science, University of California, Davis, CA, 95616, USA, \texttt{kksankar@ucdavis.edu}}}
		
\date{\today}
\begin{document}

\maketitle

\begin{flushleft}
		
		\begin{abstract}
		    We study the question of how much classical communication is needed when Alice is given a classical description of a quantum state $\ket{\psi}$ for Bob to recover any expectation value $\bra{\psi} M \ket{\psi}$ given an observable $M$ with $M$ Hermitian and $||M||_{\text{op}} \leq 1$. This task, whose study was initiated by Raz (ACM 1999) and more recently investigated by Gosset and Smolin (TQC 2019), can be thought of as a fully classical version of the pure state case of the well-known classical shadows problem in quantum learning theory. We show how the hardness of these two seemingly distinct problems are connected. 
		    
		  	We first consider the relative error version of the communication question and prove a lower bound of $\Omega(\sqrt{2^{n}}\epsilon^{-2})$ on the one-way randomized classical communication, improving upon an additive error lower bound of $\Omega(\sqrt{2^{n}})$ as shown by Gosset and Smolin. Notably, we show that this lower bound holds not only for the set of all observables but also when restricted to just the class of Pauli observables. This fact implies a $\Omega(\sqrt{2^{n}})$ versus $O(\text{poly}(n))$ separation in the compression size between the relative and additive error settings for non-adaptive Pauli classical shadows with classical memory.
		    	
		    Extending this framework, we prove randomized communication lower bounds for other relative error one-way classical communication tasks: an $\Omega(2^{n}\epsilon^{-2})$ lower bound when instead Alice is given an observable and Bob is given a quantum state and they are asked to estimate the expectation value, an $\Omega(\sqrt{n}\epsilon^{-2})$ lower bound when restricted to Paulis, and an $\Omega(\sqrt{2^{n}}\epsilon^{-2})$ lower bound when Alice and Bob are both given quantum states and asked to estimate the inner product.
		\end{abstract}
		
		\section{Introduction}
		
		What is the limit on compressing any quantum state using a classical computer? Naively, one might assume that since a generic quantum state on $n$ qubits could have $2^{n}$ non-trivial complex amplitudes, one would need to keep track of an exponential amount of information, even after being given a full classical description of a quantum state, for most interesting tasks classically. Is it possible to do any better?  \\[0.5em]
		
		We study this question in the communication complexity setting and specifically for the problem of observable estimation. This line of work was initially proposed by Raz \cite{R99}, who considered the \textit{vector in subspace} problem in the setting of  one-way randomized communication, where Alice is given a classical description of a quantum state $\ket{\psi}$ and Bob is given a projector $\Pi$ and asked to decide with high probability whether $\bra{\psi} \Pi \ket{\psi} \geq \frac{2}{3}$ or $ \bra{\psi} \Pi \ket{\psi} \leq \frac{1}{3},$ given a promise that one of the two cases is true for any particular input. Surprisingly, he showed that you only need $O(\sqrt{2^{n}})$ one-way randomized communication to solve this problem, defying the natural intuition. But was this the best you could do classically? 
		
		Later on, Gosset and Smolin \cite{GS19} considered the more general question of \textit{observable expectation estimation} for any $M$ with $M$ Hermitian and $||M||_{\text{op}} \leq 1,$ and not considering a decision promise problem but rather the question of approximating $\bra{\psi} M \ket{\psi}$ up to additive error with high probability. They gave a more time efficient compression-estimation scheme with one-way randomized communication complexity $\tilde{O}(\sqrt{2^{n}} \epsilon^{-1})$ and an almost matching lower bound of $\Omega(\sqrt{2^{n}}).$ They were able to prove this lower bound by first showing an $\Omega(\sqrt{2^{n}})$ lower bound on the vector in subspace problem, and then reducing it to observable estimation. So up to poly$(n)$ factors this was optimal in $n$ the number of qubits, essentially resolving both questions in this setting. Again this result was surprising since Alice has no information about which observable Bob gets yet can do better than just sending the entire state. These works motivate the fairly broad question:
			
		\begin{center}
		\textit{Given a classical description of a quantum state, can we characterize the limits of compression for interesting classical communication tasks like observable estimation?}
		\end{center}

		It turns out that studying this question has implications on other problems, such as the classical shadows task in quantum learning theory  \cite{AA18} \cite{HKP20} \cite{GPS22}. In the classical shadows task, we are given copies of a quantum state specified by its density matrix $\rho$ and asked to learn enough properties about it to compute $Tr(M\rho)$ for any observable $M$ up to additive error. There are usually two steps, measurement and estimation, where classical information gained from the measurement step is used in estimation to then compute some properties of interest. \\[0.5em]
		
		 Most of the complexity results in classical shadows prove bounds on the sample complexity of the task, but recently there has been interest in more fine-grained analysis of time and space efficient algorithms as well \cite{CGY24} \cite{CGZ24}. Without efficient measurement and estimation algorithms, it may never be possible to run them in practice, even with optimal sample complexity. Moreover, this problem is highly relevant to near term quantum computing, as there has been theoretical evidence that we can get a quantum learning advantage with quantum memory compared to classical memory \cite{CCHL21}. How does this relate to communication complexity? Lower bounds on the observable expectation estimation task can be thought of as putting a hard limit on the size of \textit{compressed classical descriptions} of quantum states for non-adaptive (in the sense that estimation strictly follows after all measurements are completed) algorithms that solve classical shadows.  \\[0.5em]
		
		While Gosset-Smolin tackled compressed classical descriptions in the generic observable setting, one can also consider a more restricted set of observables, like the set of Pauli operators. Aaronson for example considered the problem of shadow tomography, where a finite set of observables is actually known at the time of measurement. It can be inferred from his work that when restricted to Pauli observables, there exist $O(\text{poly}(n))$ compressed classical descriptions of quantum states \cite{AA05} \cite{AA18}. Recent work has shown that one can do even better for Pauli shadow tomography, getting \textit{rapid retrieval} with compressed descriptions of size $O(\text{poly}(n))$ and decompression algorithms running in $O(\text{poly}(n))$ time in the $O(1)$ additive error setting \cite{RGKB24}. Yet no matching lower bounds on the compression size in this setting seems to be known currently.  \\[0.5em]
		
		While the model of $O(1)$ additive error is practically useful since it matches up with expected error models on NISQ devices, a natural question one might ask is whether anything changes in the relative or multiplicative error setting, where the allowed error is much stricter. While the compression lower bound proved by Gosset-Smolin carries over, there aren't any lower bounds known to us in the case of compressed classical descriptions for relative error Pauli shadow tomography. More generally, we ask:
		
		\begin{center}
		\textit{Can we improve lower bounds on compressed classical descriptions for  classical shadows or shadow tomography under stricter assumptions on allowed error or the set of observables?}
		\end{center}
		
		\subsection{Summary of Results}
		
		We tackle these questions as follows (fixing the number of qubits to be $n$). We first improve Gosset-Smolin's $\Omega(\sqrt{2^{n}})$ one-way randomized classical communication bound to $\Omega(\sqrt{2^{n}} \epsilon^{-2})$ in \hyperlink{thm2.4}{Theorem 2.4} by reducing from the Gap-Hamming Distance problem (drawing inspiration from work on proving lower bounds for sketching quadratic forms in \cite{ACK15}) in the relative error setting, instead of the vector in subspace problem. We prove that
		 
		\begin{customthm}{2.4}
		 		The randomized one-way communication complexity of the $\epsilon-$relative error $\langle M \rangle$ problem with $\ket{\psi} \in \mathbb{C}^{2^{n}}$ is $\Omega(\sqrt{2^{n}} \epsilon^{-2})$ for $\epsilon \in (\frac{1}{2^{n/4}}, 1 ).$
		 \end{customthm}
		 
		 We then consider a restriction of the same problem to only Pauli observables in \hyperlink{thm3.1}{Theorem 3.1}  and show relative error approximations will still take $\Omega(\sqrt{2^{n}} \epsilon^{-2})$ communication as well. 
		 
		 \begin{customthm}{3.1} 
		  The randomized one-way communication complexity of the $\epsilon-$relative error $\langle M \rangle$ problem with $\ket{\psi} \in \mathbb{C}^{2^{n}}$ is $\Omega(\sqrt{2^{n}} \epsilon^{-2})$ for $\epsilon \in (\frac{1}{2^{n/4}}, 1 )$ even when $M$ is restricted to only Pauli observables.
		  \end{customthm}
		 
		 Next we relate these arguments to proving bounds on the compression size for the relative error version of the classical shadows/tomography problem in \hyperlink{thm4.1}{Theorem 4.1}. Notably, we show that relative error Pauli observable estimation with \textit{any} number of nonadaptive measurements and classical memory cannot be space efficient with respect to the compression size (polynomial in the number of qubits). That is, we have the reduction:
		 
		 \begin{customthm}{4.1} 
		 Given a pair of algorithms $(\mathscr{A}_{\text{meas}}, \mathscr{A}_{\text{est}})$ that solves $\epsilon-$relative error Classical Shadows Task over some set of observables $Obs,$ we can use them to construct an algorithm that solves the $\epsilon-$relative error $\langle M \rangle $ problem relative to that same set of observables and with the same one-way randomized communication cost. 
		\end{customthm}
	
		Additionally, we use this general framework to give lower bounds for other related communication tasks. In \hyperlink{thm5.1}{Theorem 5.1}, we consider the problem where Alice is now given a generic observable and has to compress it in order to recover the any expectation value, given Bob knows a classical description of a quantum state. We prove a lower bound of $\Omega(2^{n} \epsilon^{-2})$ on the one-way randomized communication complexity under relative error. 
		
		 \begin{customthm}{5.1}  
		The randomized one-way communication complexity of the $\epsilon-$relative error $\langle M \rangle$ observable problem across any set of observables on $n$ qubits is $\Omega(2^{n} \epsilon^{-2})$ for $\epsilon \in (\frac{1}{\sqrt{2^{n}}}, 1 ).$
		\end{customthm}	
		
		In \hyperlink{thm6.1}{Theorem 6.1}, we prove a $\Omega(\sqrt{n} \epsilon^{-2})$ lower bound on the same relative error observable compression task when restricted to Paulis. Note that there is a straightforward algorithm that gives an upper bound of $O(n).$ 
		
		\begin{customthm}{6.1} 
		The randomized one-way communication complexity of the $\epsilon-$relative error $\langle M \rangle$ observable problem when restricted to Paulis on $n$ qubits is $\Omega(\sqrt{n} \epsilon^{-2})$ for $\epsilon \in (\frac{1}{n^{1/4}}, 1 ).$
		\end{customthm}	
	
		And finally in \hyperlink{thm7.1}{Theorem 7.1}, we consider the task of relative error inner product estimation, and prove a lower bound of $\Omega(\sqrt{2^{n}} \epsilon^{-2}).$
		
		\begin{customthm}{7.1} 
				The randomized one-way communication complexity of $\epsilon-$relative error inner product estimation with $\ket{\psi} \in \mathbb{C}^{2^{n}}$ and $\ket{\phi} \in \mathbb{C}^{2^{n}}$  is $\Omega(\sqrt{2^{n}} \epsilon^{-2})$ for $\epsilon \in (\frac{1}{2^{n/4}}, 1 ).$
		\end{customthm}		
		
		\subsection{Related Results}
		\begin{itemize}
		\item Gosset and Smolin \cite{GS19} gave a lower bound of $\Omega(\sqrt{2^{n}})$ for the additive version of this problem; we study the relative error version of the problem, improving this bound and showing a new bound on Pauli observables.
		\item Andoni et al. \cite{ACK15} studied a version of the observable compression problem but only for PSD matrices and with no restriction on the norm of the vector, whereas we consider Hermitian and bounded operator norm matrices, Pauli matrices, and only vectors representable as quantum states.
		\item Estimating the inner product for real vectors $u$ and $v$ such that $||u||_{2}=||v||_{2}=1$ in the one-way randomized communication setting was considered in a recent work by Andoni et al. \cite{ABF22}, but their lower bound is $\Omega ( \epsilon^{-2} )$ for the additive error task, whereas we improve this bound in the relative error setting to $\Omega ( \sqrt{2^{n}} \epsilon^{-2} ),$ given a quantum state on $n$ qubits.
		\end{itemize}
		
		\subsection{A Comment on Precision}
		
		Since we are considering classical communication and quantum states are specified with complex amplitudes, a natural question is how we would represent these states on a classical computer, since infinite precision is computationally infeasible. \\[0.5em]
		
		We claim our results will hold (assuming we reserve $n$ for the number of qubits) with $O(n)$ bit precision for all reductions. Additionally in the communication complexity setting we are allowed unbounded computation for both individual parties, so asking for $O(1)$ precision (for example) isn't required. 
		
		\section{Preliminaries and General Case}

	   \hypertarget{def2.1}{}
		\begin{definition}[$\langle M \rangle$ Problem, Additive Error]
		Alice is given $\epsilon > 0, \ket{\psi} \in \mathbb{C}^{2^{n}},$ and $ p \in \lbrack 0, 1 \rbrack.$ Bob is given $M$ (where $||M||_{\text{op}} \leq 1$ and $M$ is Hermitian) and must return an estimate $E$ of $\bra{\psi} M \ket{\psi}$ with the condition that 
		$$\text{Pr}(|E - \bra{\psi} M \ket{\psi}| > \epsilon) \leq p $$ 
		\end{definition}		
		
	   \hypertarget{def2.2}{}
	   \begin{definition}[$\langle M \rangle$ Problem, Relative Error] 
	   Alice is given $\epsilon > 0, \ket{\psi} \in \mathbb{C}^{2^{n}},$ and $ p \in \lbrack 0, 1 \rbrack.$ Bob is given $M$ (where $||M||_{\text{op}} \leq 1$ and $M$ is Hermitian) and must return an estimate $E$ of $\bra{\psi} M \ket{\psi}$ with the condition that 
		$$\text{Pr}(|E - \bra{\psi} M \ket{\psi}| > \epsilon \cdot |\bra{\psi} M \ket{\psi}|) \leq p $$ 
		\end{definition}	
		
		We can equivalently think of relative error approximations as being bounded in the range of $(1 - \epsilon) \cdot \bra{\psi} M \ket{\psi}$ and $(1 + \epsilon) \cdot \bra{\psi} M \ket{\psi}$ with high probability.
		
		\begin{definition}[Randomized One-Way Communication Complexity, \cite{GPS22}] $R_{\delta}(f)$, the bounded-error randomized one-way communication complexity of $f,$ is the minimum number of bits that Alice must send to Bob with a public-coin protocol to compute $f$ over all possible inputs with failure probability at most $\delta.$
		\end{definition}
		
		Specifically, we will consider the standard definition of two-sided bounded-error randomized communication with failure probability $\delta = \frac{1}{3}.$		

	   \hypertarget{def2.3}{}
		\begin{definition}[Indexing Problem] Alice is given a bitstring $x \in \lbrace 0 , 1 \rbrace^{n}.$ Bob is given an index $i \in \lbrack n \rbrack =  \lbrace 1, 2, \dotsc, n \rbrace .$ Bob's goal is to estimate the bit at position $i$ in $x$ (aka $x_{i}$) given a message from Alice.
		\end{definition}
	
		\hypertarget{thm2.1}{}
		\begin{theorem}[Lower Bound on Indexing with SR]
		To solve the Indexing Problem with success probability $\frac{1}{2} + \delta$ for $\delta \in (0, \frac{1}{2} \rbrack,$ Alice needs to send Bob $\Omega (n)$ bits in the randomized one-way communication complexity model with shared randomness.
		\end{theorem}
		
		\textit{Proof.} This is a standard result in communication complexity, for example see Section 3 of \cite{JKS08}.
		
		\hypertarget{def2.4}{}
		\begin{definition}[Hamming Distance]
		Given $x, y \in \lbrace 0, 1 \rbrace^{n}$ we define $\Delta(x, y)$ to be the Hamming Distance which is the number of positions such that $x_{i} \neq y_{i}.$
		\end{definition}
		
		\hypertarget{thm2.2}{}
		\begin{theorem} 
		Given that $x$ and $y$ are bitstrings in $\lbrace 0, 1 \rbrace^{n}$, we have $\Delta(x, y)$ = \text{nnz}(x) + \text{nnz}(y) - $2\langle x, y \rangle $ where $nnz(\cdot)$ denotes the number of positions that are 1 in a bitstring and $\langle \cdot, \cdot \rangle$ denotes the standard inner product. 
		\end{theorem}
		
		\textit{Proof.} Since we know from counting the positions where the bitstrings have 1 (where $overlap$ is the set of positions $x$ and $y$ overlap and $ones_{x}/ones_{y}$ refers to the set of positions with 1 in the respective strings), we have the equation
		$$  \text{nnz}(x) + \text{nnz}(y) = \# \lbrace overlap \land ones_{x} \rbrace + \#\lbrace overlap \land ones_{y} \rbrace + \# \lbrace \lnot overlap \land ones_{x} \rbrace +  \# \lbrace \lnot overlap \land ones_{y} \rbrace $$
		and 
		$$\# \lbrace overlap \land ones_{x} \rbrace  = \# \lbrace overlap \land ones_{y} \rbrace = \langle x, y \rangle$$ and $$\# \lbrace \lnot overlap \land ones_{x} \rbrace +  \# \lbrace \lnot overlap \land ones_{y} \rbrace = \Delta(x, y).$$
		Now rewriting $\text{nnz}(x) + \text{nnz}(y) - 2\langle x, y \rangle,$ we can see that it will be equal to $\Delta(x, y).$
	
		\hypertarget{thm2.3}{}
		\begin{theorem}[Reduction of Indexing to Gap Hamming Distance with SR, Modified Version of {\cite[Lemma~5.3]{ACK15}}]
				Let $x$ be a random bitstring of length $\gamma = \frac{1}{\epsilon^{2}}$ and let $i$ be a random index in $\lbrack \gamma \rbrack.$ Choose $C\gamma$ public random bitstrings $r^{1}, \dotsc, r^{C \gamma}$ each of length $\gamma,$ where $C$ is a constant. Create $C\gamma$ length bitstrings $a, b$ as follows: 
				\begin{itemize}
				\item For each $j \in \lbrack C\gamma \rbrack,$ $a_{j} = \text{majority} \lbrace r_{k}^{j} \; | \; \text{indices } k \text{ for which } x_{k} = 1 \rbrace $
				\item For each $j \in \lbrack C\gamma \rbrack, b_{j} = r_{i}^{j}$ 
				\end{itemize}
				There is a procedure which with probability $\frac{1}{2} + \delta$ for some (fixed) constant $\delta \in (0, \frac{1}{2} \rbrack$ can determine the value of $x_{i}$ from any $d \sqrt{ C \gamma }-$additive approximation to $\Delta (a, b),$ provided $d \geq 0$ is a sufficiently small constant. 
		\end{theorem}
		
		See \hyperlink{A1}{Appendix A} for the proof of \hyperlink{thm2.3}{Theorem 2.3} and explicit values for the constants. 
		
		The goal is to show that we have a reduction between the relative error version of the expectation value problem and indexing. Formally, we have the following theorem.
		
		\hypertarget{thm2.4}{}
		\begin{theorem} 
		The randomized one-way communication complexity of the $\epsilon-$relative error $\langle M \rangle$ problem with $\ket{\psi} \in \mathbb{C}^{2^{n}}$ is $\Omega(\sqrt{2^{n}} \epsilon^{-2})$ for $\epsilon \in (\frac{1}{2^{n/4}}, 1 ).$
		\end{theorem}
		
		\textit{Proof.} Let $\gamma = \epsilon^{-2}$ and set $N' = (\sqrt{2^{n}}- \epsilon^{-2})\epsilon^{-2}.$ Suppose we consider an instance of the indexing problem with Alice receiving $z \in \lbrace 0, 1 \rbrace ^{N'}$ and Bob receiving $l \in \lbrack N' \rbrack$. Alice creates strings $z^{1}, z^{2}, \dotsc, z^{\sqrt{2^{n}} - \gamma}$ (by partitioning $z$) and $r^{1}, \dotsc, r^{C\gamma}$ by choosing random public strings. Alice then uses those to create $a^{1}, \dotsc, a^{\sqrt{2^{n}}- \gamma}, b^{1}, \dotsc, b^{\gamma}$ where each string is of length $C\gamma$ based on \hyperlink{thm2.3}{Theorem 2.3}. \\[0.5em]
		
		Now assume we have a one-way procedure for the relative error $\langle M \rangle$ problem with $N = 2^{n+q}$ with $C \leq 2^{q}$ where $q$ is also a constant. \\[0.5em]
		
		We can interpret  $a^{1}, \dotsc, a^{\sqrt{2^{n}}- \gamma}, b^{1}, \dotsc, b^{\gamma}$ as column vectors with entries in $\lbrace 0, 1 \rbrace.$ Define them as $\hat{a}^{1}, \dotsc, \hat{a}^{\sqrt{2^{n}}- \gamma},\hat{b}^{1}, \dotsc, \hat{b}^{\gamma}.$ Let Alice construct 
		$$\ket{\psi} = \frac{1}{\sqrt{D}}\begin{bmatrix} \hat{a}^{1} \\ \vdots \\ \hat{a}^{\sqrt{2^{n}}-\gamma} \\ \hat{b}^{1} \\ \vdots \\ \hat{b}^{\gamma} \\ 0 \\ \vdots \\ 0 \end{bmatrix}$$
		 where $D$ is a normalization constant dependent on the choice of $a$ and $b,$ and we have stacked column vectors on top of each other to form $\ket{\psi},$ which we can interpret as corresponding to a state in the standard basis on $\mathbb{C}^{2^{n+q}}$. We pad with zeros depending on the size of $C(\sqrt{2^{n}}- \epsilon^{-2})\epsilon^{-2} \leq 2^{n+q}.$  \\[0.5em]
		 
		 We also have Bob rewrite the index $l = i + (j-1)\cdot\gamma$ with $i \in \lbrack \gamma \rbrack$ and $j \in \lbrack \sqrt{2^{n}}- \gamma \rbrack$ to choose $M_{l}$ with the following properties. We set $M_{l}$ to have nonzero rows only corresponding to the locations of the entries of $\hat{a}^{j}$ in the resulting vector with $\frac{1}{\sqrt{2}}$ at exactly just the columns corresponding to $\hat{a}^{j}$ and $\hat{b}^{i}$ for each of the rows. It is easiest to write this in block matrix notation as follows:
		  
		  \setcounter{MaxMatrixCols}{20}
		 $$M_{l} = \begin{bmatrix}
		 	 		0 & \cdots & \cdots & \cdots & \cdots & \cdots & \cdots & \cdots  & \cdots & \cdots  & 0  \\
		 		\vdots & \vdots & \vdots & \vdots & \vdots  & \vdots   & \vdots   & \vdots   & \vdots   & \vdots & \vdots   \\
		 		 0 & \cdots & \cdots & \cdots & \cdots & \cdots & \cdots & \cdots  & \cdots & \cdots  & 0  \\
		 0 & \cdots & 0 & \frac{1}{\sqrt{2}}I_{C\gamma} & 0 & \cdots & 0 & \frac{1}{\sqrt{2}}I_{C\gamma}  & 0
		  & \cdots & 0 \\
		  0 & \cdots & \cdots & \cdots & \cdots & \cdots & \cdots & \cdots  & \cdots & \cdots  & 0  \\
		 \vdots & \vdots & \vdots & \vdots & \vdots  & \vdots   & \vdots   & \vdots   & \vdots   & \vdots  & \vdots  \\
		  0 & \cdots & \cdots & \cdots & \cdots & \cdots & \cdots & \cdots  & \cdots & \cdots & 0   \\
		 \end{bmatrix}$$
		 
		 where every entry consists of a $C \gamma \times C \gamma$ matrix, with row $j$ having the matrix $\frac{1}{\sqrt{2}}I_{C\gamma}$ at columns corresponding to $j$ and $\sqrt{2^{n}}- \gamma + i$ and the rest zeros. The resulting action on $\ket{\psi}$ will be  \\[0.5em]
		 
		 $$M_{l} \ket{\psi} = \begin{bmatrix} 0 \\ \vdots \\ 0 \\ \frac{1}{\sqrt{2D}}(\hat{a}^{j} + \hat{b}^{i}) \\ 0 \\ \vdots \\ 0 \end{bmatrix}$$

		with the non-zero entry at position $j$ in the block vector. Note that $||M_{l}||_{\text{op}} \leq 1$ since each of its non-zero rows are orthonormal. We then set $M = M_{l}^{\dagger}M_{l}$ which will be Hermitian with $||M||_{\text{op}}= ||M_{l}^{\dagger}M_{l}||_{\text{op}} \leq 1.$ $M$ can also be represented with finite precision. So we get $$\bra{\psi} M_{l}^{\dagger}M_{l} \ket{\psi} = \frac{1}{2D}||\hat{a}^{j} + \hat{b}^{i}||^{2}_{2}.$$  \\[0.5em]
		
		So Alice sends $\ket{\psi}$ to Bob as chosen before and Bob chooses $M = M_{l}^{\dagger}M_{l}$ as from before. Finally Bob can compute an estimate $E$ such that with high probability it is between $(1- \epsilon) \frac{1}{2D} ||\hat{a}^{j} + \hat{b}^{i}||^{2}_{2}$ and $(1+ \epsilon) \frac{1}{2D} ||\hat{a}^{j} + \hat{b}^{i}||^{2}_{2}.$ After multiplying by $2D$ (which Alice can send as well using $o(\sqrt{2^{n}} \epsilon^{-2})$ bits), we proceed exactly as in the proof of Theorem 5.2 from \cite{ACK15}.  \\[0.5em]
		
		That is, we now have a relative error approximation to $||\hat{a}^{j} + \hat{b}^{i}||^{2}_{2}.$ From here, we can turn this into a $d_{1}\epsilon^{-1}$ additive error estimate to $\langle \hat{a}^{j}, \hat{b}^{i} \rangle,$ since
		$$ 2 \langle \hat{a}^{j}, \hat{b}^{i} \rangle = ||\hat{a}^{j} + \hat{b}^{i}||^{2}_{2} -  ||\hat{a}^{j}||^{2}_{2} - ||\hat{b}^{i}||^{2}_{2} $$
		 
		And by using this definition and noting  \hyperlink{thm2.2}{Theorem 2.2}, we can get a $d \sqrt{C\gamma}$ additive approximation of $\Delta(a^{j}, b^{i}).$ Note that we can send $||\hat{a}^{j}||^{2}_{2} \text{ and } nnz(a^{j})$ for all $j \in \lbrack \sqrt{2^{n}} - \gamma \rbrack$ using $o(\sqrt{2^{n}} \epsilon^{-2})$ bits. With shared randomness, we have access to $||\hat{b}^{i}||^{2}_{2} \text{ and } nnz(b^{i})$ for all $i \in \lbrack \gamma \rbrack.$ Finally, if we assume this relative error procedure errs with probability $\delta^{'} = \frac{1}{3}$ and combining this with \hyperlink{thm2.3}{Theorem 2.3}, we have the $\langle M \rangle$ algorithm solves indexing with probability
		
		\begin{align*}
		1 - (\delta^{'} + (1-\delta^{'})(\frac{1}{2} - \delta)) &\geq 1 - (\delta^{'} + (\frac{1}{2} - \delta)) \\
		&\geq 1 - \delta^{'} - \frac{1}{2} + \delta \\
		&\geq \frac{1}{2} + (\delta - \delta^{'})
		\end{align*}
		
		Since $\delta > \frac{1}{3}$ (since $\delta + \frac{1}{2} > \frac{5}{6}$) this allows us to solve the indexing problem of size $N' = (\sqrt{2^{n}}- \epsilon^{-2})\epsilon^{-2}$ given a relative error $\langle M \rangle$ procedure on $n+q$ qubits with probability $\geq \frac{1}{2} + (\delta - \delta^{'}).$ Applying \hyperlink{thm2.1}{Theorem 2.1} will give us a lower bound of $\Omega(\sqrt{2^{n}}\epsilon^{-2}).$
		
		\section{Pauli Observables}
		
		What if we restricted our observables $M$ to only Pauli observables? Formally, we consider elements of $P_{n}$ the Pauli group on $n$ qubits. We claim the lower bound still holds, that is:
		 
		\hypertarget{thm3.1}{}
		 \begin{theorem} 
		 The randomized one-way communication complexity of the $\epsilon-$relative error $\langle M \rangle$ problem with $\ket{\psi} \in \mathbb{C}^{2^{n}}$ is $\Omega(\sqrt{2^{n}} \epsilon^{-2})$ for $\epsilon \in (\frac{1}{2^{n/4}}, 1 )$ even when $M$ is restricted to only Pauli observables.
		 \end{theorem}
		 
		\textit{Proof.} Again, set $\gamma = \epsilon^{-2}$ and set $N^{'} = (\sqrt{2^{n}} - \gamma) \gamma.$ We again consider an instance of the indexing problem with Alice receiving $z \in \lbrace 0, 1 \rbrace ^{N'}$ and Bob receiving $l \in \lbrack N' \rbrack$. Alice creates strings $z^{1}, z^{2}, \dotsc, z^{\sqrt{2^{n}}- \gamma}$ (by partitioning $z$) and $r^{1}, \dotsc, r^{C\gamma}$ by choosing random public strings. Alice then uses those to create $a^{1}, \dotsc, a^{\sqrt{2^{n}}- \gamma}, b^{1}, \dotsc, b^{\gamma}$ where each string is of length $C\gamma$ based on \hyperlink{thm2.3}{Theorem 2.3}. \\[0.5em]
		 
		 Treating these strings as vectors in $\lbrace 0, 1 \rbrace$ (define them as $\hat{a}^{1}, \dotsc, \hat{a}^{\sqrt{2^{n}}- \gamma},\hat{b}^{1}, \dotsc, \hat{b}^{\gamma}$), Alice now computes every possible combination of $$||\hat{a}^{j} + \hat{b}^{i}||_{2}^{2}$$ for $j \in \lbrack \sqrt{2^{n}}- \gamma \rbrack$ and $i \in \lbrack \gamma \rbrack.$ There will be $(\sqrt{2^{n}} - \gamma) \gamma = N^{'}$ possible values of this form which we use to recover the index. \\[0.5em]
		 
		 Now we use the following fact:
		 
		 \hypertarget{fact3.1}{}
		  \begin{fact} 
		  Consider the set of Pauli operators on $n$ qubits that are formed from the tensor product of only $I$ or $Z$ (where $I$ and $Z$ are the standard Pauli single qubit operators). These operators (which can be thought of as a string of length $n$ composed of elements in $\lbrace I, Z \rbrace$) are linearly independent over $\mathbb{C}.$
		  \end{fact}
		  
		  Now consider the matrix representation of these operators (formed only from a tensor product of $I$ or $Z$) over the standard basis. Fixing some number of qubits $n,$ define $row(P)$ to be exactly the row vector representation of the diagonal entries from any $P \in P_{n}$ such that $P$ is exactly a tensor product of $I$ and $Z$ (Note: this is different from $vec(P)$, the vector representation of the \textit{entire} matrix including zeros). So $row(P)$ will be a $1 \times 2^{n}$ vector with values in $\lbrace \pm 1 \rbrace.$ \\[0.5em]
		  
		  Stacking these row vectors together, we can construct a $2^{n} \times 2^{n}$ matrix that we call $Diag(P_{n})$. Writing this explicitly, we define (with respect to the standard computational basis)
		  
		  $$ Diag(P_{n}) := \begin{bmatrix}
		  row(P^{1}) \\
		  row(P^{2}) \\
		  \vdots \\
		  row(P^{2^{n}})
		  \end{bmatrix}$$
		  
		  where $P^{1}, \dotsc P^{2^{n}}$ correspond exactly to unique elements of $P_{n}$ that are formed from a tensor product of $I$ and $Z.$ The order in which we label them does not matter for the following lemma, but we will discuss where the labeling comes into play later. 
		  
		  We claim the following:
		  
		 \hypertarget{lemma3.2}{}
		  \begin{lemma} 
		  $Diag(P_{n})$ is an invertible matrix over $\mathbb{C}$. Additionally, $Diag(P_{n})^{-1} = \frac{1}{2^{n}} Diag(P_{n})^{\dagger}$
		  \end{lemma}	
		  
		  \textit{Proof.} This is true as a direct consequence of \hyperlink{fact3.1}{Fact 3.1}. Since we know those operators are diagonal matrices in the computational basis and linearly independent over $\mathbb{C}$ we must have $Diag(P_{n})$ has linearly independent rows and is invertible. \\[0.5em]
		  
		  We also can conclude from \hyperlink{fact3.1}{Fact 3.1} and the orthogonality of the Pauli basis that  $row(P^{k}) row(P^{k})^{\dagger} = 2^{n}$ for all $k$ and $row(P^{k}) row(P^{k^{'}})^{\dagger} = 0$ for all $k \neq k^{'}.$ So we must have:
		  
		  \begin{align*}
		  Diag(P_{n})\frac{1}{2^{n}} Diag(P_{n})^{\dagger} &= \frac{1}{2^{n}} \begin{bmatrix}
		  		  row(P^{1}) \\
		  		  row(P^{2}) \\
		  		  \vdots \\
		  		  row(P^{2^{n}}) 
		  		  \end{bmatrix} \begin{bmatrix}
		  		  		  		  row(P^{1})^{\dagger} & row(P^{2})^{\dagger} & \cdots & row(P^{2^{n}})^{\dagger}
		  		  		  		  \end{bmatrix} \\
		  		  		  		 &= I_{2^{n}}
		  \end{align*}
		  
		 Now construct the following vector of stacked norms (assume we fix a basis, here we assume the standard computational basis again):
		 $$ \begin{bmatrix} ||\hat{a}^{1} + \hat{b}^{1}||_{2}^{2} \\ \vdots \\ ||\hat{a}^{1} + \hat{b}^{\gamma}||_{2}^{2} \\ ||\hat{a}^{2} + \hat{b}^{1}||_{2}^{2}  \\ \vdots \\ ||\hat{a}^{2} + \hat{b}^{\gamma}||_{2}^{2} \\ \vdots \\ ||\hat{a}^{\sqrt{2^{n}} - \gamma} + \hat{b}^{1}||_{2}^{2}  \\ \vdots \\ ||\hat{a}^{\sqrt{2^{n}} - \gamma} + \hat{b}^{\gamma}||_{2}^{2} \\ 0 \\ \vdots \\ 0 \end{bmatrix}.$$
		 
		 where we pad the vector with zeros depending on the length. Note  $N^{'} = (\sqrt{2^{n}} - \gamma) \gamma \leq 2^{n}$ by construction. We then can compute a vector 
		 
		 $$v = Diag(P_{n})^{-1} \begin{bmatrix} ||\hat{a}^{1} + \hat{b}^{1}||_{2}^{2} \\ \vdots \\ ||\hat{a}^{1} + \hat{b}^{\gamma}||_{2}^{2} \\ ||\hat{a}^{2} + \hat{b}^{1}||_{2}^{2}  \\ \vdots \\ ||\hat{a}^{2} + \hat{b}^{\gamma}||_{2}^{2} \\ \vdots \\ ||\hat{a}^{\sqrt{2^{n}} - \gamma} + \hat{b}^{1}||_{2}^{2}  \\ \vdots \\ ||\hat{a}^{\sqrt{2^{n}} - \gamma} + \hat{b}^{\gamma}||_{2}^{2} \\ 0 \\ \vdots \\ 0 \end{bmatrix} $$
		 
		 A note on the size of $v.$ We can bound the size of $||a^{j} + b^{i}||^{2}_{2}$ by $\gamma \leq 2^{n/2}.$ Putting this together with \hyperlink{lemma3.2}{Lemma 3.2}, we can conclude that the range of any entry in $v$ will be $\lbrack - \gamma, \gamma \rbrack$ up to $2^{-n}$ increments. So we can represent any entry in this vector using $O(n)$ bit precision, and its norm can be represented using $O(\log (2^{n} \gamma^{2})) \leq O(n + \log \gamma)$ bits. So $v$ can be represented with finite precision in space $O(2^{n}).$
		 
		 We will have Alice send the vector
		 
		 $$\ket{v^{'}} = \frac{1}{\sqrt{D}} \begin{bmatrix} v \\ 1 \\ \vdots \\ 1 \end{bmatrix}$$
		 
		 where we add $2^{n}$ additional 1s and $D = 2^{n} + \sum_{i} v_{i}^{2}.$ Note $v$ is a $2^{n} \times 1$ vector (we concatenate it with 1s). We can interpret $\ket{v^{'}}$ as a $n+1$ qubit state, where $v$ corresponds to elements with $\ket{0}$ as the last qubit, the 1s correspond to $\ket{1}$ with the last qubit. Again by a similar argument to the paragraph above, we can represent $\ket{v^{'}}$ using finite precision, so we can pass it as an input to our $\langle M \rangle$ algorithm. Additionally, $D$ can be represented in space $O(n + \log \gamma) = o(\sqrt{2^{n}} \epsilon^{-2}).$ \\[0.5em]
		 
		 Alternatively we could write this state in bra-ket notation as
		 
		 $$ \ket{v^{'}} = \frac{1}{\sqrt{C^{'}}} (\ket{v}\ket{0} + \sum_{y \in \lbrace 0, 1 \rbrace^{n}} \ket{y}\ket{1}) $$
		 
		 where $\ket{v}$ is the normalized vector $v$ written in the computational basis and $C^{'} = 1 + 2^{n}.$
		 
		 So what does Bob do?  Depending on which index $l = i + (j-1)\cdot\gamma$ he cares about, he chooses the corresponding Pauli observable $P_{l}$ composed of only $Z$ or $I,$ which he derives from the corresponding row in $Diag(P_{n}).$ Note the ordering $P^{1}, \dotsc, P^{2^{n}}$ can be fixed in advance, or Alice and Bob can use the public random strings to derive the ordering, so no communication is needed from Alice for Bob to make this decision. Finally he tensors $P_{l}$ with $X.$  \\[0.5em]
		 
		 So computing $\bra{v^{'}} (P_{l} \otimes X ) \ket{v^{'}},$ we will get 
		 \begin{align*}
		 \bra{v^{'}} (P_{l} \otimes X ) \ket{v^{'}} &= \frac{1}{\sqrt{C^{'}}} \bra{v^{'}} (P_{l} \ket{v} \ket{1} + \sum_{y \in \lbrace 0, 1 \rbrace^{n}} P_{l}\ket{y}\ket{0}) \\
		 &= \frac{1}{C^{'}} ( \sum_{y \in \lbrace 0, 1 \rbrace ^{n}} \bra{y} P_{l} \ket{v} \bra{1}\ket{1} +  \sum_{y \in \lbrace 0, 1 \rbrace ^{n}}  \bra{v} P_{l} \ket{y} \bra{0}\ket{0}) 
		 \end{align*}
		 
		 where we have expanded the vectors and propagated the final $X$. \\[0.5em]
		 
		  Now we note the fact that the sum of the entries in $P_{l}v$ is equal to $||\hat{a}^{j} + \hat{b}^{i}||_{2}^{2}$ (by our choice of defining the matrix-vector product $Diag(P_{n}) v $ to be equal to the vector of stacked norms). So we get  
		 
		 \begin{align*}
		 \bra{v^{'}} (P_{l} \otimes X ) \ket{v^{'}} &= \frac{1}{D} ( ||\hat{a}^{j} + \hat{b}^{i}||_{2}^{2} + ||\hat{a}^{j} + \hat{b}^{i}||_{2}^{2})  \\
		 &= \frac{2}{D}( ||\hat{a}^{j} + \hat{b}^{i}||_{2}^{2}).
		 \end{align*}
		 
		 So to summarize, Alice computes $\ket{v^{'}}$ and sends it to Bob, who sets $M = P_{l} \otimes X$ which is a Pauli matrix. Finally Bob can compute the estimate $E$ such that with high probability it is between $(1- \epsilon) \frac{2}{D} ||\hat{a}^{j} + \hat{b}^{i}||^{2}_{2}$ and $(1+ \epsilon) \frac{2}{D} ||\hat{a}^{j} + \hat{b}^{i}||^{2}_{2}.$ From here, we scale the estimate by $D/2$ (which Alice can send as well using $o(\sqrt{2^{n}} \epsilon^{-2})$ bits) and proceed again exactly as in the previous proof.
		 
		 This allows us to solve the indexing problem of size $N' = (\sqrt{2^{n}}- \epsilon^{-2})\epsilon^{-2}$ given a relative error $\langle M \rangle$ procedure (restricted to Paulis) with $n+1$ qubits, which gives us a lower bound of $\Omega(\sqrt{2^{n}}\epsilon^{-2})$ from \hyperlink{thm2.1}{Theorem 2.1}. 
		 
	\section{Connection to Classical Shadows}
	
	We are also able to relate these one-way communication lower bounds to the relative error classical shadows problem. The following definitions are all adapted from \cite{GPS22},
	
	\hypertarget{def4.1}{}
	\begin{definition}[Observables] Define a set of observables as
	$$Obs := \lbrace O \in \mathbb{C}^{2^{n} \times 2^{n}} \mid O = O^{\dagger}, ||O||_{\text{op}} \leq 1 \rbrace$$
	\end{definition}
	
	\hypertarget{def4.2}{}
	\begin{definition}[Classical Shadows Task]
	The Classical Shadows Task consists of two separate phases - a measurement phase and an observable estimation phase - which are completed by two separate (randomized) algorithms $\mathscr{A}_{meas}$ and $\mathscr{A}_{est},$ respectively. 
	
	\textbf{Measurement:} $\mathscr{A}_{meas}: \rho^{\otimes s} \rightarrow \lbrace 0, 1 \rbrace^{*} $
	\begin{itemize}
	\item Input: $s$ copies of a state $\rho \in \mathbb{C}^{2^{n} \times 2^{n}}$
	\item Output: A bitstring
	\end{itemize}
	
	\textbf{Estimation:} $\mathscr{A}_{est} : Obs \times \lbrace 0, 1 \rbrace^{*}  \rightarrow \mathbb{R} $
	\begin{itemize}
	\item Input: Observable $O \in Obs$ and a classical shadow
	\item Output: Estimate $E \in \mathbb{R}$
	\end{itemize}
	\end{definition}
	
	Restating from \cite{GPS22}, we note that the input to the measurement algorithm is quantum (the state $\rho^{\otimes s}$) and the output is classical (the classical shadow). The output is computed from measuring the input state with some POVM (with arbitrary post-processing). \\[0.5em]
	
	For the purposes of our results, we will consider a relative error validness criteria. That is, we will say that $\mathscr{A}_{\text{meas}}$ and $\mathscr{A}_{\text{est}}$ constitute a valid protocol for the classical shadows task if their estimate for the expectation of the observable $E := \mathscr{A}_{\text{meas}}(O, \mathscr{A}_{\text{est}}(\rho^{\otimes s}))$ is such that
	$$ |\text{Tr}(O\rho) - E | \leq \epsilon \cdot | \text{Tr}(O\rho) | $$
	
	with probability at least $1 - \delta$ over the randomness in $\mathscr{A}_{\text{meas}}$ and $\mathscr{A}_{\text{est}}.$
	
	\hypertarget{def4.3}{}
	\begin{definition}[Pauli Shadow Tomography]  
	The Classical Shadows Task where $Obs = P_{n},$ the set of all Paulis on $n$ qubits. 
	\end{definition}
	
	\hypertarget{thm4.1}{}
	\begin{theorem} 
	Given a pair of algorithms $(\mathscr{A}_{\text{meas}}, \mathscr{A}_{\text{est}})$ that solves $\epsilon-$relative error Classical Shadows Task over some set of observables $Obs,$ we can use them to construct an algorithm that solves the $\epsilon-$relative error $\langle M \rangle $ problem relative to that same set of observables and with the same one-way randomized communication cost. 
	 \end{theorem}	
	 
	 \textit{Proof.} Here we will need to consider a more general notion of a reduction, since we can't use a communication complexity reduction directly between a classical input and quantum input problem. To do this, we will need to consider one more problem which we define as follows
	 
	 \hypertarget{def4.4}{}
	 \begin{definition}[Classical Input Classical Shadows]
	 We define this problem to be exactly the same as the Classical Shadows problem, but instead of being given copies of a quantum state $\rho$ as input to $\mathscr{A}_{\text{meas}},$ we are given a singly copy of a full classical description of $\rho.$
	 \end{definition}
	 
	 Any $(\mathscr{A}_{\text{meas}}, \mathscr{A}_{\text{est}})$ for Classical Shadows can be converted into a fully classical set of measurement and estimation algorithms that solves Classical Input Classical Shadows -- we can simply use the classical description of $\rho$ to simulate the quantum state and sample accordingly based on the measurements specified in $\mathscr{A}_{\text{meas}}.$ The estimation algorithm, which is already known to be classical, is the same (apply  $\mathscr{A}_{\text{est}}$ as a black box). \\[0.5em]
	 
	 Finally since Classical Input Classical Shadows is an entirely classical problem, we can easily reduce the $\langle M \rangle$ problem to it. Alice can create the matrix $\ket{\psi} \bra{\psi}$ and Bob can pass his observable to the estimation algorithm. So the theorem is proved. \\[0.5em]
	 
	 From this reduction, we can conclude that any learning algorithm that claims to solve the $\epsilon$-relative error Classical Shadows problem on $n$ qubits must output a classical compressed representation that is of size $\Omega(\sqrt{2^{n}} \epsilon^{-2})$ from the lower bound we derived in \hyperlink{thm2.4}{Theorem 2.4}. Note that any lower bound on a protocol with public randomness implies a lower bound on one with private randomness, which allows our claim to hold. \\[0.5em]
	 
	 Similarily, the $\Omega(\sqrt{2^{n}} \epsilon^{-2})$ lower bound also holds for the restriction of the observables to the set of all Paulis from \hyperlink{thm3.1}{Theorem 3.1}. \\[0.5em] 
	 
	 This tells us a space complexity separation between the additive error and relative error settings for Pauli tomography, as it known how to generate a $O(\text{poly}(n))-$sized classical compressed representation in the additive error setting (for example, see \cite{RGKB24}), yet we show this cannot be possible in the relative error setting. 
		 		 
		 \section{Communicating Observables to recover $\langle M \rangle$}
		 
		 Now we consider a related one-way communication task, namely we swap the inputs that are given to Alice and Bob (so now Alice gets an observable and Bob gets a classical description of a quantum state). Formally, we define
		 
		\hypertarget{def5.1}{}
		 \begin{definition}[$\langle M \rangle$ Observable Problem, Relative Error]
		 		Alice is given $\epsilon > 0, p \in \lbrack 0, 1 \rbrack,$ and $M$ a classical description of an observable on $n$ qubits (with $M$ Hermitian, $||M||_{\text{op}} \leq 1$).  Bob is given $\ket{\psi} \in \mathbb{C}^{2^{n}}$ a classical description of a quantum state and must return an estimate $E$ of $\bra{\psi} M \ket{\psi}$ with the condition that 
		 		$$\text{Pr}(|E - \bra{\psi} M \ket{\psi}| > \epsilon \cdot |\bra{\psi} M \ket{\psi}|) \leq p $$ 
		 \end{definition}
		 
		 Depending on the set of observables we care about, we will have different representations for $M.$ For example, if we let $M$ be a generic observable we will need to represent it as a $2^{n} \times 2^{n}$ matrix, whereas if $M$ is restricted to a Pauli, we will consider representing it as a $O(n)$ length classical bitstring instead.  
		 
		 We start by proving a lower bound on the most general case:
		 
		\hypertarget{thm5.1}{}
		 \begin{theorem} 
		 The randomized one-way communication complexity of the $\epsilon-$relative error $\langle M \rangle$ observable problem across any set of observables on $n$ qubits is $\Omega(2^{n} \epsilon^{-2})$ for $\epsilon \in (\frac{1}{\sqrt{2^{n}}}, 1 ).$
		 \end{theorem}		 
		
	\textit{Proof.} Set $\gamma = \epsilon^{-2}$ and set $N^{'} = (2^{n} - \gamma) \gamma.$ We again consider an instance of the indexing problem with Alice receiving $z \in \lbrace 0, 1 \rbrace ^{N'}$ and Bob receiving $l \in \lbrack N' \rbrack$. Alice creates strings $z^{1}, z^{2}, \dotsc, z^{2^{n} - \gamma}$ (by partitioning $z$) and $r^{1}, \dotsc, r^{C\gamma}$ by choosing random public strings. Alice then uses those to create $a^{1}, \dotsc, a^{2^{n} - \gamma}, b^{1}, \dotsc, b^{\gamma}$ where each string is of length $C\gamma$ based on \hyperlink{thm2.3}{Theorem 2.3}. \\[0.5em]
	
	Now construct the following matrix  of size $C\gamma \times 2^{n}$ interpreting the strings as column vectors as we did previously,
	
	$$ M = \begin{bmatrix} \hat{a}^{1} &  \hat{a}^{2} & \cdots &  \hat{a}^{2^{n}-\gamma} & \hat{b}^{1} & \cdots & \hat{b}^{\gamma} \end{bmatrix} $$
	
	Alice sends $\frac{1}{||M^{\dagger}M||_{\text{op}}} M^{\dagger}M$ to Bob and Bob uses a pure state with exactly two nonzero elements (corresponding to the columns of $M$ that he cares about) to compute a relative error estimate of
	$$\frac{1}{2||M^{\dagger}M||_\text{op}} ||\hat{a}^{j} + \hat{b}^{i}||^{2}_{2} $$
	
	From here we use the same argument as before to recover the index after scaling the estimate. Note if $M$ is all zeros, we do not normalize. We also can send $||M^{\dagger} M||_{\text{op}}$ using $o(2^{n} \epsilon^{-2})$ bits. 
	
	This allows us to solve the indexing problem of size $N' = (2^{n}- \epsilon^{-2})\epsilon^{-2}$ given a relative error $\langle M \rangle$ observable procedure on $n$ qubits, which gives us a lower bound of $\Omega(2^{n}\epsilon^{-2})$ from \hyperlink{thm2.1}{Theorem 2.1}.
	
	\section{Communicating Observables: Restriction to Paulis}
	
	Now suppose Alice has an observable that is a Pauli and wants to send enough information about it to recover the expectation value. Clearly there is a naive deterministic algorithm which uses $O(n)$ bits. But is this the best that we can do? The following lower bound could shed some light on that question.
	
	\hypertarget{thm6.1}{}
	\begin{theorem} 
	The randomized one-way communication complexity of the $\epsilon-$relative error $\langle M \rangle$ observable problem when restricted to Paulis on $n$ qubits is $\Omega(\sqrt{n} \epsilon^{-2})$ for $\epsilon \in (\frac{1}{n^{1/4}}, 1 ).$
	\end{theorem}		
	
	\textit{Proof.}  Set $\gamma = \epsilon^{-2}$ and set $N^{'} = (\sqrt{n} - \gamma) \gamma.$ We again consider an instance of the indexing problem with Alice receiving $z \in \lbrace 0, 1 \rbrace ^{N'}$ and Bob receiving $l \in \lbrack N' \rbrack$. Alice creates strings $z^{1}, z^{2}, \dotsc, z^{\sqrt{n} - \gamma}$ (by partitioning $z$) and $r^{1}, \dotsc, r^{C\gamma}$ by choosing random public strings. Alice then uses those to create $a^{1}, \dotsc, a^{\sqrt{n} - \gamma}, b^{1}, \dotsc, b^{\gamma}$ where each string is of length $C\gamma$ based on \hyperlink{thm2.3}{Theorem 2.3}. \\[0.5em]
	
	We can imagine concatenating $a^{1}, \dotsc, a^{\sqrt{n} - \gamma}, b^{1}, \dotsc, b^{\gamma}$ into one long bitstring of size $C\gamma\sqrt{n} .$ Based on this bitstring, Alice sends
	$$ P = Z^{a^{1}_{1}} \otimes Z^{a^{1}_{2}} \otimes \cdots \otimes Z^{a^{1}_{C\gamma}} \otimes Z^{a^{2}_{1}} \otimes \cdots \otimes Z^{a^{2}_{C\gamma}} \otimes \cdots \otimes Z^{b^{\gamma}_{1}} \otimes \cdots \otimes Z^{b^{\gamma}_{C\gamma}} \otimes Z$$
	
	Bob can recover a scaled version of $\Delta(a^{j}, b^{i})$ by choosing his vectors carefully. All he needs to do is use the following subset state on $n+ q +1 = n^{'}+1$ (unnormalized state for now and note $C \leq 2^{q}$)
	$$ \sum_{x_{k} \in S} \ket{x_{k}0} $$
	where $S$ consists of exactly $C\gamma$ bitstrings, where each bitstring $x_{k} \in \lbrace 0, 1 \rbrace^{n+q}$ has exactly 2 1s at indices corresponding to $a_{k}^{j}$ and $b_{k}^{i}$ for all $k \in \lbrack C\gamma \rbrack$ and zeros elsewhere. 
	
	He then creates the state
	
	$$ \ket{\psi^{'}} = \frac{1}{\sqrt{2C\gamma}} (\ket{\psi} + \sqrt{C\gamma} \ket{0^{n^{'}}1})$$
	
	where $\ket{\psi}$ is the un-normalized subset state from before (we abuse notation here). 
	
	So computing 
	\begin{align*}
	\bra{\psi^{'}} P \ket{\psi^{'}}  &= \frac{1}{2C\gamma}( \bra{\psi}P\ket{\psi} + C\gamma \bra{0^{n^{'}}1}P\ket{0^{n^{'}}1})\\
	&=\frac{1}{2C\gamma} ( \#_{\text{same}} ( a^{j}, b^{i} ) - \#_{\text{different}}( a^{j}, b^{i} )- C\gamma) \\
	&= -\frac{2}{2C\gamma} \cdot \#_{\text{different}} ( a^{j}, b^{i} )
	\end{align*}
		
	where we note that $\#_{\text{same}}/\#_{\text{diff}}$ counts the number of indices that have the same/different bits for the input strings, and 
	$$\bra{\psi}P\ket{\psi} =  \#_{\text{same}} ( a^{j}, b^{i} ) - \#_{\text{different}}( a^{j}, b^{i} )$$
	by our choice of $P$ and $\ket{\psi}.$ We also use the fact that 
	$$  \#_{\text{same}} ( a^{j}, b^{i} ) + \#_{\text{different}}( a^{j}, b^{i} ) = C \gamma$$
	for any choice of $i \in \lbrack \gamma \rbrack$ and $j \in \lbrack \sqrt{n} - \gamma \rbrack$ to simplify the expression above.
	
	From here we have estimated a scaled version of $\#_{\text{different}} ( a^{j}, b^{i} ) = \Delta(a^{j}, b^{i})$ which we can again use the argument from \cite{ACK15} and the previous proofs to achieve the reduction. We can also send $C \gamma$ using $o(\sqrt{n}\epsilon^{-2})$ bits. 
	
	This allows us to solve the indexing problem of size $N' = (\sqrt{n} - \epsilon^{-2})\epsilon^{-2}$ given a relative error $\langle M \rangle$ observable procedure (restricted to Paulis) on $n+q+1$ qubits, which gives us a lower bound of $\Omega(\sqrt{n}\epsilon^{-2})$ from \hyperlink{thm2.1}{Theorem 2.1}.
		
	\section{Relative Error Inner Product Estimation in Communication Setting}
	
		\hypertarget{def7.1}{}
			 \begin{definition}[Inner Product Estimation, Relative Error]
			 		Alice is given $\epsilon > 0, p \in \lbrack 0, 1 \rbrack,$ and $\ket{\psi} \in \mathbb{C}^{2^{n}}$ a classical description of a quantum state on $n$ qubits.  Bob is given $\ket{\phi} \in \mathbb{C}^{2^{n}}$ a classical description of a quantum state on $n$ qubits and must return an estimate $E$ of $ \mid \bra{\phi} \ket{\psi} \mid $ with the condition that 
			 		$$\text{Pr}\lbrack (1 - \epsilon) \cdot \mid \bra{\phi} \ket{\psi} \mid \leq E \leq (1+ \epsilon) \cdot \mid \bra{\phi} \ket{\psi} \mid \rbrack \geq 1-p$$ 
			 \end{definition}
			 
			We claim the following:
			
			\hypertarget{thm7.1}{}
			\begin{theorem} 
				The randomized one-way communication complexity of $\epsilon-$relative error inner product estimation with $\ket{\psi} \in \mathbb{C}^{2^{n}}$ and $\ket{\phi} \in \mathbb{C}^{2^{n}}$  is $\Omega(\sqrt{2^{n}} \epsilon^{-2})$ for $\epsilon \in (\frac{1}{2^{n/4}}, 1 ).$
			\end{theorem}		
			
			\textit{Proof.}  Set $\gamma = \epsilon^{-2}$ and set $N^{'} = (\sqrt{2^{n}} - \gamma) \gamma.$ We again consider an instance of the indexing problem with Alice receiving $z \in \lbrace 0, 1 \rbrace ^{N'}$ and Bob receiving $l \in \lbrack N' \rbrack$. Alice creates strings $z^{1}, z^{2}, \dotsc, z^{\sqrt{2^{n}} - \gamma}$ (by partitioning $z$) and $r^{1}, \dotsc, r^{C\gamma}$ by choosing random public strings. Alice then uses those to create $a^{1}, \dotsc, a^{\sqrt{2^{n}} - \gamma}, b^{1}, \dotsc, b^{\gamma}$ where each string is of length $\gamma$ based on \hyperlink{thm2.3}{Theorem 2.3}. \\[0.5em]
		
			We can interpret  $a^{1}, \dotsc, a^{\sqrt{2^{n}}- \gamma}, b^{1}, \dotsc, b^{\gamma}$ as column vectors with entries in $\lbrace 0, 1 \rbrace.$ Define them as $\hat{a}^{1}, \dotsc, \hat{a}^{\sqrt{2^{n}}- \gamma},\hat{b}^{1}, \dotsc, \hat{b}^{\gamma}.$ Note there exists $q$ such that $C \leq 2^{q}.$ Let Alice construct a vector on $n+q$ qubits consisting of these stacked column vectors (assuming standard computational basis),
			$$\ket{\psi} = \frac{1}{\sqrt{D}}\begin{bmatrix}
			\hat{a}^{1} \\
			\hat{a}^{2} \\
			\vdots \\
			\hat{a}^{\sqrt{2^{n}} - \gamma} \\
			0 \\ 
			\vdots \\
			0
			\end{bmatrix}$$
			
			Bob likewise constructs the vector
			$$\ket{\phi} = \frac{1}{\sqrt{D^{'}}}\begin{bmatrix}
						0 \\
						\vdots \\
						0 \\
						\hat{b}^{i} \\
						0 \\ 
						\vdots \\
						0
			\end{bmatrix}$$
			
			where the location of $\hat{b}^{i}$ depends on which $\Delta(a^{j}, b^{i})$ we want to compute corresponding to $l = i + (j-1) \cdot \gamma$. Specifically, treating $\ket{\phi}$ as a block vector, we will have $\hat{b}^{i}$ at position $j.$ Given a way to estimate $(1+\epsilon) \mid \bra{\psi} \ket{\phi} \mid,$ we claim we are able to estimate $\Delta(a^{j}, b^{i})$ to the desired approximation as well, since there exists a constant $c^{'}$ such that $c^{'}\epsilon \cdot \mid \bra{\psi} \ket{\phi} \mid \leq c^{'} \epsilon \gamma \leq c^{'} / \epsilon.$ So since we have the desired additive error approximation to the inner product, this naturally gives us the desired one for the hamming distance, and the argument goes through as in the previous reductions.
			
		This allows us to solve the indexing problem of size $N' = (\sqrt{2^{n}} - \epsilon^{-2})\epsilon^{-2}$ given a relative error inner product estimation procedure on $n+q$ qubits, which gives us a lower bound of $\Omega(\sqrt{2^{n}}\epsilon^{-2})$ from \hyperlink{thm2.1}{Theorem 2.1}.			
			
	\section{Acknowledgements}
	
	I thank Isaac Kim for the helpful discussions and for providing extensive feedback on drafts of these results and Daniel Grier for pointing me to \cite{ACK15}.
			
	 \bibliographystyle{alphaurl}
	 \bibliography{QCLB6}
		
	\appendix
		
		\section{Proof of Theorem 2.3}
		
		This proof is not original and is based on \cite{JKS08}, \cite{R15}, and  \cite{W07}. Additionally, we will make our constants explicit. Restating the theorem, we have: \\[0.5em]
		
		\hypertarget{A1}{}
		 \begin{customthm}{2.3}[Reduction of Indexing to Gap Hamming Distance with SR, Modified Version of {\cite[Lemma~5.3]{ACK15}}]
				Let $x$ be a random bitstring of length $\gamma = \frac{1}{\epsilon^{2}}$ and let $i$ be a random index in $\lbrack \gamma \rbrack.$ Choose $C\gamma$ public random bitstrings $r^{1}, \dotsc, r^{C \gamma}$ each of length $\gamma,$ where $C$ is a constant. Create $C\gamma$ length bitstrings $a, b$ as follows: 
				\begin{itemize}
				\item For each $j \in \lbrack C\gamma \rbrack,$ $a_{j} = \text{majority} \lbrace r_{k}^{j} \; | \; \text{indices } k \text{ for which } x_{k} = 1 \rbrace $
				\item For each $j \in \lbrack C\gamma \rbrack, b_{j} = r_{i}^{j}$ 
				\end{itemize}
				There is a procedure which with probability $\frac{1}{2} + \delta$ for some (fixed) constant $\delta \in (0, \frac{1}{2} \rbrack$ can determine the value of $x_{i}$ from any $d \sqrt{ C \gamma}-$additive approximation to $\Delta (a, b),$ provided $d \geq 0$ is a sufficiently small constant. 
		\end{customthm}
		
		\textit{Proof.} We will set $C = 9 \cdot \frac{1}{c^{2}}$ where $0 < c < \sqrt{\frac{2}{\pi}}.$ This will allow us to solve the indexing problem with probability $ \geq 1 - e^{-2} > \frac{5}{6}$ given a $d\sqrt{ \Delta (a, b)}$ additive approximation with $0 \leq d < \frac{1}{2}.$ The justification for these constants will be shown through the proof. Note that this analysis on the constants is probably not optimal but suffices for our results. \\[0.5em]
		
		Assume $r \sim \mu$ where $\mu$ is the uniform distribution on $\gamma$ length bitstrings. Define 
		$$a_{0} = \text{majority} \lbrace r_{k} \; | \; \text{indices } k \text{ for which } x_{k} = 1 \rbrace$$  
		
		where $a_{0} \in \lbrace 0, 1 \rbrace.$ Then we claim
		
		$$\text{Pr} \lbrack a_{0} \neq r_{i} \rbrack =  \begin{cases}
		 \geq \frac{1}{2} & \text{ if } x_{i} = 0 \\
		  \leq \frac{1}{2} - \frac{c}{\sqrt{\gamma}} &  \text{ if } x_{i} = 1
		\end{cases}$$
		
		To prove this, first let $\gamma$ be odd. Consider
		$$ S =  \lbrace r_{k} \; | \; \text{indices } k \text{ for which } x_{k} = 1, k \neq i \rbrace $$ 
		which corresponds to the set of bits in $r$ whose corresponding bits (at the same position) in $x$ are 1, excluding at position $i.$ Now we have two cases depending on the value of $x_{i}:$
		
		\begin{itemize}
		\item Case 1: $x_{i} = 0.$ If this is the case, then $a_{0}$ must just reduce to maj$(S).$ But maj$(S)$ is independent of $r_{i},$ so we have
			\begin{align*}
				Pr \lbrack a_{0} \neq r_{i} | x_{i} = 0 \rbrack &= Pr \lbrack \text{maj}(S) \neq r_{i} | x_{i} = 0 \rbrack \\
				& \geq \frac{1}{2}
			\end{align*}
		\item Case 2: $x_{i} = 1.$ We use the following lemma from \cite{W07} which states:
		\begin{lemma}[{\cite[Lemma~61]{W07}}]
		Let $m$ be a sufficiently large odd integer. There is a constant $0 < c < \sqrt{\frac{2}{\pi}}$ such that for i.i.d. Bernoulli(1/2) random variables $B_{1}, \dotsc, B_{m},$ for any $i, 1 \leq i \leq m,$
		$$ Pr \lbrack MAJ(B_{1}, \dotsc, B_{m}) = 1 \mid B_{i} = 1 \rbrack > \frac{1}{2} + \frac{c}{\sqrt{m}},$$
		where $MAJ(B_{1}, \dotsc, B_{m}) = 1$ iff the majority of the $B_{i}$ are 1. 
		\end{lemma}
		So applying this here we have, 
		\begin{align*}
		Pr \lbrack a_{0} = 1 \mid r_{i} = 1, x_{i} = 1 \rbrack > \frac{1}{2} + \frac{c}{\sqrt{
	\gamma}}
		\end{align*}
		which implies 
		$$Pr \lbrack a_{0} = 0 \mid r_{i} = 1, x_{i} = 1 \rbrack  < \frac{1}{2} - \frac{c}{\sqrt{\gamma}}$$
		and using symmetry and noting $r_{i}$ is 0 or 1 with probability $1/2$, we can conclude
		$$Pr \lbrack a_{0} \neq r_{i} | x_{i} = 1 \rbrack  < \frac{1}{2} - \frac{c}{\sqrt{\gamma}}$$
		\end{itemize}

		\begin{fact}[Hoeffding's Inequality]
		
		Let $X_{1}, X_{2}, \dotsc, X_{N}$ be $N$ i.i.d. 0-1 random variables and $X = \sum_{k=1}^{N} X_{k}.$ Then for $\nu \geq 0,$
		$$\text{Pr} \lbrack X - \mathbb{E} \lbrack X \rbrack > \nu \rbrack \leq e^{-2 \nu^{2}/N} 	 \text{ and }  	\text{Pr} \lbrack X - \mathbb{E} \lbrack X \rbrack < -\nu \rbrack \leq e^{-2 \nu^{2}/N} $$
		
		\end{fact}
		
		We choose $X_{k}$ to correspond to the indicator:
		
		$$ X_{k} =  \begin{cases}
				 1 & a_{k} \neq b_{k} \\
				  0 & a_{k} = b_{k}
				\end{cases}$$
				
		Now consider $X = \sum_{k=1}^{N} X_{k}.$ Note that this exactly encodes the random variable corresponding to $\Delta(a, b),$ where the choice of $b$ is dependent on a fixed $i \in \lbrack \gamma \rbrack.$ We can apply the Hoeffding bound by setting $N = 9 \cdot \gamma/c^{2}$ and $\nu = \sqrt{N}.$ This will break down into two cases:
		
		\begin{itemize}
		\item Suppose $x_{i} = 0.$ Then we have $\mathbb{E}\lbrack \Delta(a, b)  \rbrack = \mathbb{E} \lbrack X \rbrack \geq \frac{N}{2}$ and
		\begin{align*}
		Pr \lbrack \Delta(a, b)  < \frac{N}{2} - \sqrt{N} \rbrack &=  Pr \lbrack X - \frac{N}{2} < -\sqrt{N}\rbrack \\
		&\leq Pr \lbrack X - \mathbb{E} \lbrack X \rbrack < -\nu \rbrack \\
		&\leq e^{-2}
		\end{align*}
		\item Suppose $x_{i} = 1.$ Then we have $\mathbb{E} \lbrack \Delta(a, b) \rbrack = \mathbb{E} \lbrack X \rbrack \leq \frac{N}{2} - c \frac{N}{\sqrt{\gamma}}$ and
		
		\begin{align*}
		\frac{N}{2} - c\frac{N}{\sqrt{\gamma}} + \sqrt{N} &= \frac{N}{2} - c \cdot  9 \cdot \frac{\sqrt{ \gamma}}{c^{2}} + 3 \cdot \frac{\sqrt{\gamma}}{c} \\
		&= 	\frac{N}{2} - 6 \cdot \frac{\sqrt{\gamma}}{c^{2}} \\
		&= \frac{N}{2} - 2 \sqrt{N}
		\end{align*}
		
		which implies
		
		\begin{align*}
		Pr \lbrack \Delta(a, b)  > \frac{N}{2} - c\frac{N}{\sqrt{\gamma}} + \sqrt{N}  \rbrack &= Pr \lbrack X - (\frac{N}{2} - c\frac{N}{\sqrt{\gamma}}) > \sqrt{N}  \rbrack \\
		&\leq Pr \lbrack X - \mathbb{E} \lbrack X \rbrack > \nu \rbrack  \\
		&\leq e^{-2}
		\end{align*}
		\end{itemize}

		Finally we can conclude that for any $i \in \lbrack \gamma \rbrack$ (which paramaterizes our choice of $b$),

		 $$\begin{cases} 
		  Pr \lbrack \Delta(a, b) \geq \frac{N}{2} -\sqrt{N}  \rbrack \geq 1-e^{-2} > \frac{5}{6} & \text{if }x_{i} = 0 \\ 
		Pr \lbrack \Delta(a, b) \leq \frac{N}{2} - 2\sqrt{N}  \rbrack \geq 1-e^{-2} > \frac{5}{6} & \text{if } x_{i} = 1
		 \end{cases}$$
		 
		 So knowing an estimate $E = \Delta(a, b) \pm d \sqrt{N}$ for $0 \leq d < \frac{1}{2}$ will allow us to discern the two cases with high probability since the gap between the two cases is $\sqrt{N}$.  \\[0.5em]
		 
		 So all we need to do is test if our estimate E is greater than or less than $\frac{N}{2} - \frac{3}{2} \sqrt{N}$ and output $x_{i} = 0$ in the first case and $x_{i} = 1$ in the second case. This allows us to solve the indexing problem with probability greater than $1 - e^{-2} > \frac{5}{6}.$
		
\end{flushleft}
\end{document}